# Enhanced superconductivity of YBCO interfaces: origin of high critical temperature in layered superconductors

S.S. Tinchev

Institute of Electronics, Bulgarian Academy of Sciences, Tzarigradsko Chaussee 72, Sofia 1784, Bulgaria

E-mail: stinchev@ie.bas.bg

Superconducting transition temperatures $T_c$ of the YBCO film surface and of the YBCO film/substrate interface were measured inductively. It was observed that the interface- $T_c$ is always higher then the surface - $T_c$. However deposition of silver over-layer enhances the superconducting transition temperatures. This observation was confirmed by four-point resistance measurements. In the annealed YBCO/Ag bilayers magnetic properties of the interface were observed. We believe that such phenomena are a common feature of layered systems. This two dimensional structure reminds the layered microstructure of the high-temperature superconductors and one can suppose that covering of the superconducting layer by non-superconducting layer is a condition for obtaining high critical temperatures in general.

## 1. Introduction

In a superconductor below its transition temperature ($T_c$) the current is carried by pairs of electrons (Cooper pairs). The mechanism by which two negatively charged electrons are bound together is by their interaction with the vibrations of the underlying lattice: one electron in the pair polarizes the lattice by attracting the nuclei towards it, into which a second electron is attracted.

Obviously such polarization should be stronger at the surface of the superconducting material in comparison within the bulk of the superconductor simply because the surface atoms in the lattice are weakly coupled together. In other word the surface of a superconductor should possess better superconducting properties (higher $T_c$) then inside the superconductor. Covering the surface of superconductor by other material should also change its the superconducting properties because thus vibration of the underlying lattice is influenced.

Indeed 1964 Ginzburg predicted theoretically existence of interface superconductivity [1-3]. In his letters he has pointed out that two dimensional (surface) superconductivity, as well as surface ferromagnetism must arise as general phenomena of ordering in two dimensions. He also suggested that applying dielectric or mono-molecular layers on the metal surface should create attraction between electrons and thus surface superconductivity must arise.

To our knowledge no one has found difference between superconducting properties of the surface and inside superconductor. However, effects similar to interface superconductivity were already observed in conventional (low-$T_c$) superconductors. In 1965 Rühl [4] found that the transition temperature $T_c$ of different classical superconducting thin metal films (like Sn, Pb,

Al, or In) depends on the existence of a thin oxide overlayer. He observed that this overlayer can enhance the superconducting transition temperature in some cases - Al, In and Tl or to reduce the $T_c$ in Sn or Pb. In 1967 Naugle [5] observed a change in the transition temperature of Ge coated Sn and Tl films. As compared with the pure metals the critical temperature $T_c$ for Sn is reduced, while for Tl is increased. After that a lot of work has been done in this field, but the reason for this effect remained unexplained. In some cases oscillations in the superconducting transition temperature $T_c$ of superconducting films with a coating ($Nb-SiO_2$, Mo-C, Pb-Sb, Al-SiO etc.) have been observed [6-9] as a function of the thickness of the coating.

In high-$T_c$ superconductors effects of enhancement of superconducting transition temperature were observed in bulk materials as well in thin films. In [10] it was found that doping with Au enhances $T_c$ of $YBa_2Cu_3O_7$ ceramic sample by about 1.5 K. Similar for $YBa_2O_{7-\delta}$/Ag composition [11] a superconducting transition temperature slightly higher (~ 91 K) then for pure $YBa_2Cu_3O_7$ (~ 90.5 K) was measured.

1990 we observed a similar effect in the high-$T_c$ thin film bilayer system $YBa_2Cu_3O_{7-x}$/Ag [12-14]. It was found that the deposition of a silver overlayer on the $YBa_2Cu_3O_{7-x}$ thin films enhances its superconducting transition temperature. But the papers [12, 13] were not accepted for publication and remained unpublished. Only the paper [14] presented on SQUID'91 conference was published. Later on similar observation can be found indirectly in many other publications.

Recently an enhancement of superconducting transition temperature $T_c$ was observed in bilayer systems. In [15] a deposition of thin heavily overdoped (metallic) $La_{1.65}Sr_{0.35}CuO_4$ layer was found to increase $T_c$ of the under laying $La_{2-x}Sr_xCuO_4$ films. An enhancement of $T_c$ up to 50% was measured and it was concluded that the enhanced superconductivity occurs at the interface between the layers.

Two months later similar observation was published [16] for superconductivity in bilayers consisting of an insulator ($La_2CuO_4$) and metal ($La_{1.55}Sr_{0.45}CuO_4$) neither of which is superconducting in isolation. In the bilayers, however, $T_c$ of ~ 15K or ~ 30 K depending on the layering sequence was measured. Again it was concluded that the enhanced superconductivity is an interface effect. Actually an enhancement of $T_c$ in this bilayer system was observed previously [17] but the essential role of the interfaces was not recognized at that time. In the last year, however, the interface superconductivity attracted more attention; see for example [18-20], especially after the superconductivity at ~ 200 mK was observed at $LaAlO_3$ – $SrTiO_3$ interface [21].

We have no doubt that it is the same effect we have observed earlier. Nevertheless, many fundamental aspects remain unresolved, including the understanding of its mechanism. To our opinion this is a general effect and that is the reason why now, after 19 years we feel us encouraged to try again to publish some of our old results.

**2. Expermental.**

In our experiments the superconducting transition temperature of the films was measured inductively. Inductive measurement of $T_c$ is often used to study the electromagnetic properties of superconductors, in most cases as a two-coil technique at low frequency. It is well known that even the spatial variation of the critical temperature in a sample can be obtained from an inductive $T_c$ transition measurement [22]. However, to our knowledge this technique was not used to measure and compare the critical temperature of both sides of a H$T_c$ superconducting film.

We use an eddy current method at RF frequencies (about 20-30 MHz) [23]. This method is contactless and nondestructive. To the author's knowledge it is the only method, which can be used for measurement of the electrical and magnetic properties of the substrate/film interface, where no direct access is possible. As illustrated in figure 1, this method is based on the modification of the resonant frequency of a parallel LC circuit. A spiral tank circuit coil is pressed (at a distance of 0.5 mm) against the film surface (position S) or against the substrate (position I). In the both cases eddy-currents are induced in the film. The magnitude of these screening currents is a function of the temperature-dependent film conductivity. Thus any change of sample conductivity will produce a shift in the resonance frequency of the tank circuit. The same effect can occur if the film susceptibility changes.

During the measurement, actually the tank-circuit resonance frequency is recorded as a function of temperature. The resonance frequency is measured by the instrumentation shown on figure 1. A 78 KHz signal of small amplitude modulates the frequency of the voltage controlled oscillator (VCO). If the resonance frequency is changed because of sample conductivity (or susceptibility) changes, a DC voltage appears at the lock-in output, which causes the VCO frequency to follow the resonance-frequency changes. The output signal of our measuring system is the voltage controlling the VCO frequency. This is very convenient because this voltage can be fed directly to an X-Y recorder or to an analog-digital converter. We defined the transition temperature to be the temperature at which the derivative is maximal. The temperature was measured by a platinum thin film thermometer.

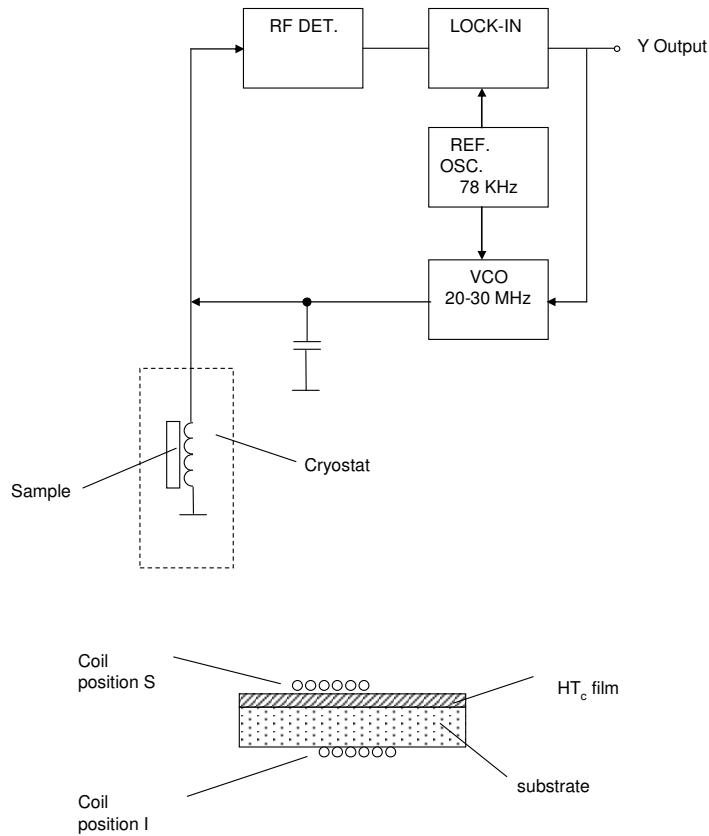

**Figure 1.** Instrumentation for eddy-current measurements of the high- $T_c$ thin films and the location of the coil over the film or below the substrate. VCO is a voltage controlled oscillator.

This method provides a useful possibility to test quickly the properties of the superconducting films coated with normal films. The most interesting property of this measurement system, however, is the capability to measure both sides of the films by placing the coil on the top of the film or below the substrate. In this way one can distinguish between the properties of the surface of film and properties of the interface film - substrate. Of course this would be possible if the films are sufficiently thick (thicker then the superconducting penetration depth, which is about 140 nm for YBCO [24]). Because the coil inductance is sensitive to the magnetic susceptibility of the sample too, it is also possible to investigate the magnetic properties of the interface as will be show below.

We also used standard 4 - point resistance measurements in order to confirm the results from the eddy current system. In some of the 4-point measurements only part of the high-$T_c$ films was covered by Ag and voltage contacts were placed on different positions to distinguish between the $T_c$ of the covered and uncovered high-$T_c$ films – figure 2.

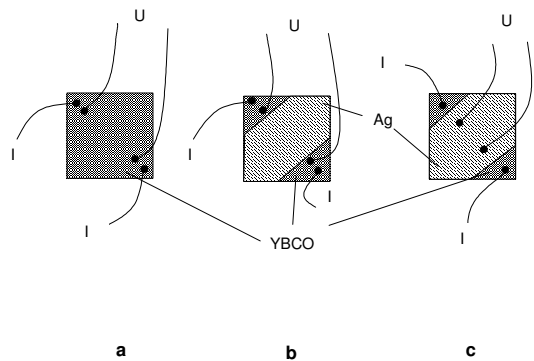

**Figure 2**. Different positions of the contacts in the 4-point resistance measurements: a) before Ag deposition b) all 4-contacts placed on the YBCO c) voltage contacts on the Ag.

## 3. Results

Figure 3 shows a typical result for the superconducting transition of 300 nm $YBa_2Cu_3O_{7-x}$ film measured before deposition of a 50 nm silver overlayer. Here the derivatives of both curves are also shown. Above the superconducting transition both curves coincided quite well. Below the $T_c$ the output voltage change of the system is smaller for the position of the coil below the substrate because simply the distance to the film is bigger in this case. It is obvious that both sides of the $YBa_2Cu_3O_{7-x}$ films have different $T_c$. The difference is small (about 0.25 K), but quite measurable. In all cases the $T_c$ of the interface is higher then the $T_c$ of the film surface probably enhanced by the presence of the substrate.

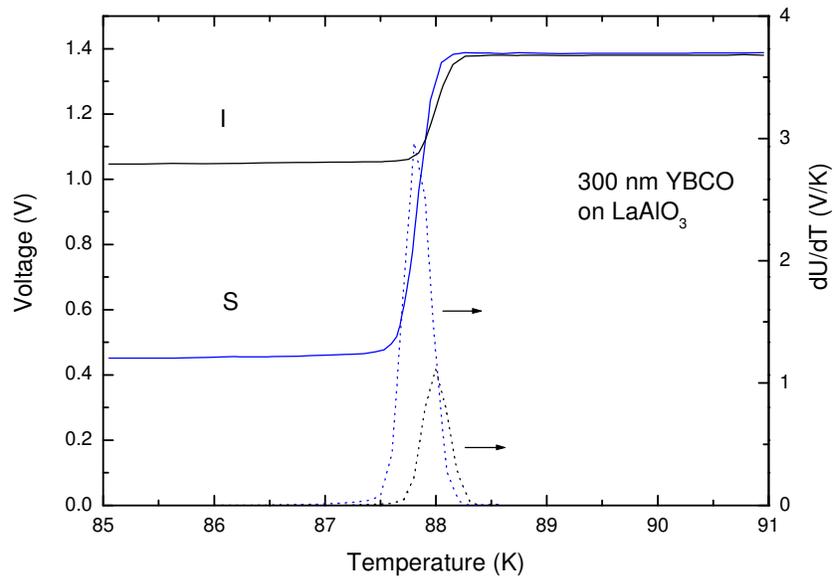

**Figure 3**. Superconducting transition of 300 nm YBa$_2$Cu$_3$O$_{7-x}$ thin film before silver deposition measured from both sides of the film (s – from the surface and i – from the interface) by the eddy current method. The derivatives of the both curves are also shown. Note the different superconducting transition temperatures for the measurements from the both film sides.

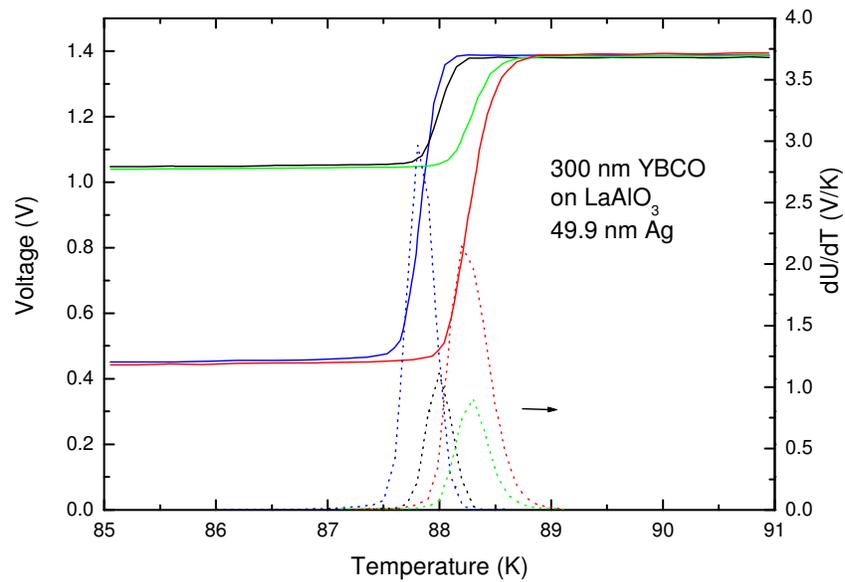

**Figure 4**. Effect of silver over-layer on superconducting properties of a 300 nm - YBa$_2$Cu$_3$O$_{7-x}$ film recorded before and after deposition of the 50 nm - Ag over-layer.

Effect of deposition of 50 nm Ag on the top of the same film is shown in figure 4. The superconducting transitions of both sides of the film are increased. Remarkable is that the enhancement of the surface-$T_c$ is higher and that after the Ag deposition the $T_c$ of the both sides of the film coincide. These observations could be explained in different ways. One can speculate for example that the original surface of the YBCO film is deteriorated because ex-situ treatment and Ag-deposition, therefore its virgin-$T_c$ is smaller. Another possible explanation seems more plausible for us. In the same way as silver enhances $T_c$ of the YBCO film surface, so could the substrate influence the film-substrate interface. In other words we observe an interface (surface) enhanced superconductivity on the interface side by the substrate and on the surface by the silver, just as predicted by Ginzburg.

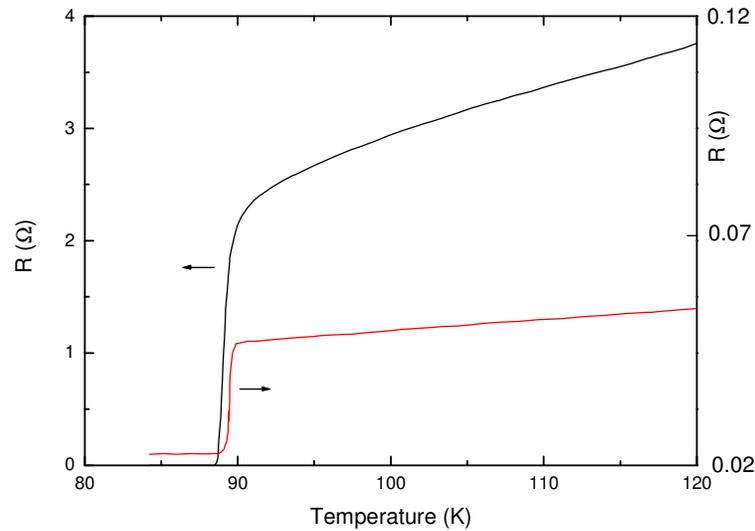

**Figure 5**. A four-point DC resistance measurement of the sample before (left curve) and after silver deposition (right curve).

Figure 5 shows result of 4-point measurements. In this first experiments silver over-layer 300 nm thick was deposited on the whole $YBa_2Cu_3O_{7-x}$ surface. All four contacts were placed as usually over the silver layer by means of a silver paint. For comparison the R(T) dependence for the same film before silver deposition is also shown. One can note enhancement of the critical temperature as well as a reduction of the transition width. The typical enhancement of $T_c$ because of silver deposition was about 0.5 K. To exclude any possible inaccuracies, for example owing to the temperature measurement of the sample, we carried out two types of other experiments using different locations of the contacts on the sample surface.

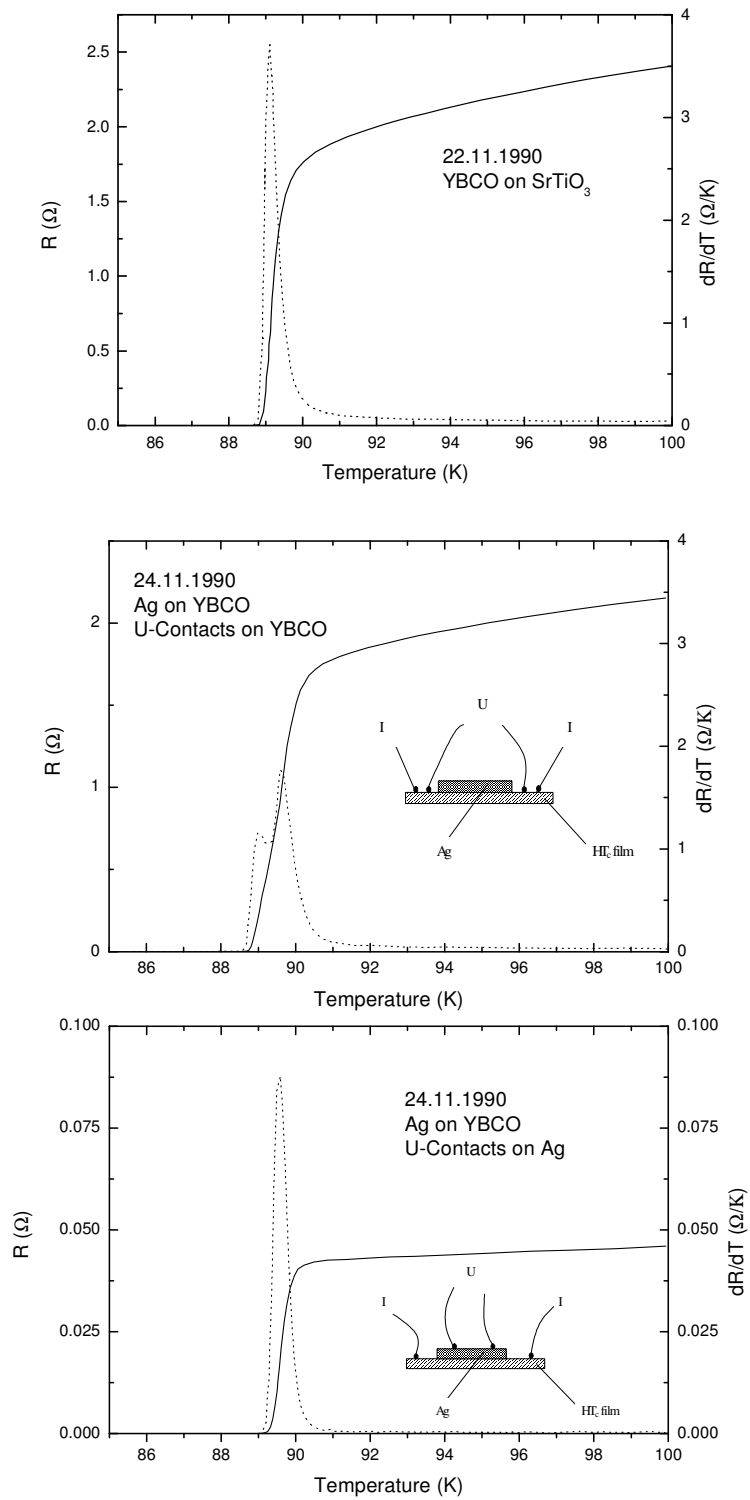

**Figure 6**. R(T) and their derivatives for YBCO film before silver deposition a) and after that b) and c) for different positions of the voltage contacts, see the insets.

In both cases the silver over-layer was deposited only over a part of the $YBa_2Cu_3O_{7-x}$ film (figure 2b, 2c). On the remained silver-free surface of the $YBa_2Cu_3O_{7-x}$ film, different contacts for the four-point resistance measurement have been placed. In the first measurements all four contacts were placed there (figure 2b). The recorded curve is shown on figure 6b. One can see that the superconducting transition consists of two parts having different slope. This can be very well recognized from the derivative $dR/dT$ depicted also in figure 6b, where two well separated peaks exist. The first one coincides exactly with the peak before Ag deposition as one can see in Fig. 6a. The second peak is obviously due to the enhanced $T_c$ of the silver covered $YBa_2Cu_3O_{7-x}$ film. To prove this suggestion in the last experiment the voltage contacts were moved over to the silver layer as is shown in figure 2c. In this case only one peak of the derivative is observed exactly at the same temperature, where the second peak was placed in figure 6b. Therefore there is no doubt that the transition temperature is increased because of the Ag deposition, and in figure 2b (all four-contacts placed outside Ag) we measured two serially connected superconductors $YBa_2Cu_3O_{7-x}$ and $YBa_2Cu_3O_{7-x}$ /Ag with different $T_c$.

Further the $T_c$ dependence on the thickness of the silver coating was investigated. In this experiment five $YBa_2Cu_3O_{7-x}$ films were coated with 20, 50, 100, 200 and 300 nm silver layers. Another set of $GdBa_2Cu_3O_{7-x}$ films was prepared in a similar way. Figure 7 gives the change of the transition temperature $T_c$ of $YBa_2Cu_3O_{7-x}$ and $GdBa_2Cu_3O_{7-x}$ films before and after Ag deposition as a function of the thickness $d_{Ag}$ of the silver coating. It is seen that for $YBa_2Cu_3O_{7-x}$ the transition temperature after the silver deposition is always higher than before Ag deposition. In $GdBa_2Cu_3O_{7-x}$ films the $T_c$ first decreases and then enhances. In both films, however, a clear oscillation of the critical temperature can be observed. The position of the maximum of $T_c$ lies in both films at a thickness of the Ag film of about 200 nm. Obviously the observed phenomena is more complicated then as expected. Besides the oscillating behavior on the silver thickness, there is also dependence on the nature of the high-$T_c$ film. However, similar observations have been made already in 1965 by Rühl [4].

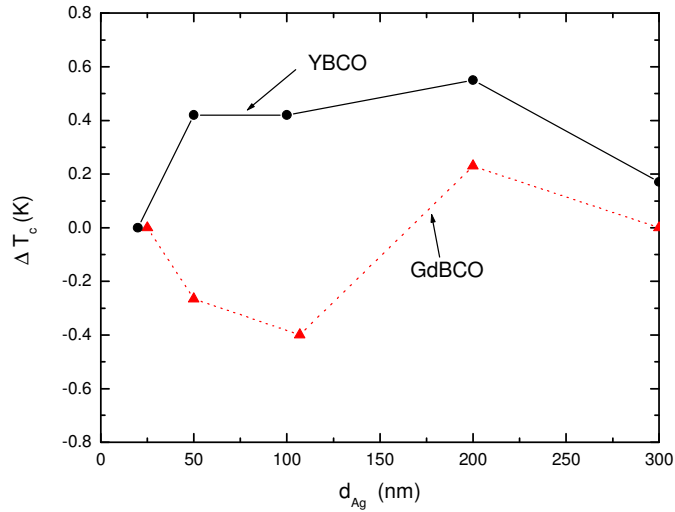

**Figure 7**: The change of transition temperature of 300 nm thick high temperature superconducting films: $YBa_2Cu_3O_{7-x}$ (circles) and $GdBa_2Cu_3O_{7-x}$ films (triangles) as a function of the thickness of the silver coating. Solid lines are drawn as guides to the eye.

Another interesting phenomenon was observed in annealed YBCO-Ag bilayers. In this experiment a 1 µm $YBa_2Cu_3O_{7-x}$ film on $LaAlO_3$ substrate was covered with 300 nm silver and the sample was annealed at 600°C for 30 minutes in oxygen. Figure 8 shows the result of the eddy current measurements of the YBCO-Ag interface. Presented here is, however, the VCO frequency change (the change of the resonant frequency of the tank circuit, see figure 1) with the sample temperature. We observed a slightly reduced $T_c$ of the bilayer and a second transition at 160 K. This transition is very sharp and is probably an antiferromagnetic transition, because below the transition the resonant frequency of the tank circuit is increased. Reduced oxygen content of the YBCO surface can be the reason for this behavior because $T_N$ (the Neel temperature) depends on the oxygen content and moves from $T_c$ to higher temperatures if oxygen content is reduced. At room temperature one can observe clear ferromagnetic properties because the resonant frequency is lower then the resonance frequency without sample (dashed line in figure 8). Therefore we concluded that YBCO/Ag interface possess magnetic properties.

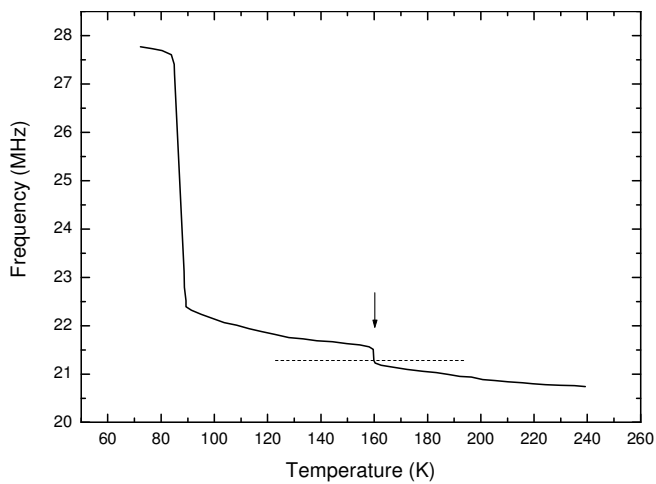

**Figure 8**. Eddy-current measurement of the $YBa_2Cu_3O_7$-Ag interface for a 1 µm thick $YBa_2Cu_3O_7$ film with 300 nm silver over-layer after 30 minutes annealing at 600°C. The arrow marks the antiferromagnetic transition at 160 K. The dashed line represented the resonance frequency without sample. The lower resonance frequency for $T \geq 160K$ indicates a ferromagnetic property of the interface.

**4. Discussion**
In the past 19 years we have found many direct and indirect confirmation of our results presented in this paper. Here I would like to mention only some of them. In [24] superconducting / normal metal bilayers using YBCO films and Ag, $PrBa_2Cu_3O_{7-\delta}$ or $(Y_{0.4}Pr_{0.6})Ba_2Cu_3O_{7-\delta}$ as normal material were investigated. In all cases it was demonstrated that the SN-bilayes showed a stronger screening compared with the superconducting film with its N-layer removed by wet chemical etching or by dry etching. The measured enhancement of the screening for the bilayers was between 7 and 20 % over the entire temperature range compared with bare YBCO film.

In [25] a system of granular Pb film with overlayer of Ag was investigated. It was observed that the Pb film which was originally insulating becomes superconducting with the addition of Ag. And what is more with Ag increasing the $T_c$ increases first and then begins to decrease similar to our observations.

The effect of amorphous $Nb_xSi_{1-x}$ on the superconducting transition temperature of ultrathin Nb films was investigated in [26]. With varied x one could move across the metal-insulator transition at x=11.5%. It was clearly demonstrated that only for insulating Nb-Si composites, $T_c$ as a function of Nb-Si thickness shows an enhancement.

In conclusion we observed that the surface and the interface of high-$T_c$ thin film with the substrate can have different $T_c$. The interface-$T_c$ is always higher then the surface-$T_c$ probably because of the influence of the substrate. Surface $T_c$ is changed (enhanced) by deposition of a silver over-layer.

In the annealed YBCO-Ag system a magnetic transition and magnetic properties of the interface were observed. Similar effect: enhancement of the Curie temperature of Fe films coated with Ag, Pd or O exists also in the magnetism [27]. Regardless of the quite different electronic structure of Ag, Pd or O, they all increase the Curie temperature by a similar amount. Therefore we believe that also in this case (in analogy to interface-superconductivity) magnetism can be induced at the interface between the otherwise nonmagnetic materials. Recently two references [28, 29] were found, where the existence of the magnetic layer at the Ag/LaCeCuO interface was also observed.

We believe that such phenomena are a common feature of layered systems. This two dimensional bi-layer structure reminds the layered microstructure of the high-temperature superconductors and one can suppose that enhancement of the interface critical temperatures by covering of the superconducting layer by non-superconducting layer is the origin of high critical temperatures in general.

**Acknowledgments**
The experiments described in the present paper were carried out during the author work at the Forschungsgesellschaft für Informationstechnik in Bad Salzdetfurth, Germany. I would like to thank Prof. J.H. Hinken for his continuous support and Mr. A. Baranyak for the high-$T_c$ films fabrication.

ELECTRICAL AND MAGNETIC PROPERTIES OF THE Y-Ba-Cu-O INTERFACES

S.S. Tinchev,

Forschungsgeselschaft für Informationtechnik,

P.O. Box 1147, D-3202 Bad Salzdetfurth, Germany

Abstract: Eddy-current technique was used for investigation of the electrical and magnetic properties of the Y-Ba-Cu-O interfaces. It was found that the interface Y-Ba-Cu-O/substrate has better superconducting properties than the top surface of the Y-Ba-Cu-O film. Deposition of a thin layer of silver without annealing improves the electrical properties of the Y-Ba-Cu-O thin film surface. Long-time annealing of the layer Y-Ba-Cu-O/Ag was found to deteriorate both the interfaces because of diffusion of the Ag into the Y-Ba-Cu-O film. If a low resistance contact to the Y-Ba-Cu-O film is required, rapid thermal annealing (flash lamp or laser) can be used. Antiferromagnetic ordering at the Y-Ba-Cu-O/Ag interface, obviously connected with the oxygen content of the interface, was observed.

**Figure 9**. Page 1 of the unpublished paper [12] submitted to J. Appl. Phys. on February 10, 1991.